\documentstyle[seceq,epsf,wrapft]{ptptex}
\newcommand{\bm}[1]{\mbox{\boldmath$#1$}}

\title{Novel Order Parameter to Characterize Valence-Bond-Solid States}
\author{Masaaki {\sc Nakamura}$^1$ and Synge {\sc Todo}$^{2,3}$}
\inst{$^1$Department of Applied Physics, Faculty of Science, Science
University of Tokyo, Kagurazaka, Shinjuku-ku, Tokyo 162-8601\\
$^2$Theoretische Physik, Eidgen\"ossische Technische Hochschule,
CH-8093 Z\"urich, Switzerland\\ $^3$Institute for Solid State Physics,
University of Tokyo, Kashiwa 277-8581}
\recdate{\today}
\abst{
 We propose an order parameter to characterize valence-bond-solid (VBS)
 states in quantum spin chains, given by the ground-state expectation
 value of a unitary operator appearing in the Lieb-Schultz-Mattis
 argument. We show that the order parameter changes the sign according
 to the configuration of the valence bonds. This allows us to determine
 the phase transition point in between different VBS states
 accurately. We demonstrate this theory in the bond-alternating
 Heisenberg chain and in the frustrated spin ladder.
}
\begin{document}
\setlength{\baselineskip}{.46cm}

\maketitle

\section{Introduction}
To study Haldane's conjecture \cite{Haldane}, Affleck, Kennedy, Lieb and
Tasaki proposed the valence-bond-solid (VBS) state, and showed that the
ground state of the spin $S=1$ Heisenberg chain is described
approximately by the VBS state \cite{Affleck-K-L-T}.  Since the spin
configurations of the VBS state show the hidden antiferromagnetic order,
den Nijs and Rommelse proposed the string order parameter (SOP) to
characterize the $S=1$ Haldane phase \cite{Nijs-R},
\begin{equation}
  {\cal{O}}^{\alpha}_{\rm string} = -\lim_{|k-l|\rightarrow\infty}
   \langle\Psi_0| S^{\alpha}_{k}
   \exp\left[{\rm i}\pi\sum_{j=k+1}^{l-1} S^{\alpha}_{j}\right]
   S^{\alpha}_{l}|\Psi_0\rangle,
   \label{eqn:string}
\end{equation}
where $\alpha= x,y,z$ and $|\Psi_0\rangle$ means the ground state. Thus
this order parameter enables us to detect the VBS state {\it
indirectly}. The SOP was generalized to $S>1$ cases by Oshikawa
\cite{Oshikawa}.

On the other hand, Affleck and Lieb studied the Haldane's conjecture by
the Lieb-Schultz-Mattis (LSM) type argument \cite{Affleck-L}.  However,
relation between the LSM argument and the VBS picture including the SOP
has not been fully understood.  In this paper, we show that the
ground-state expectation value of a unitary operator appearing in the
LSM argument, for a spin chain with length $L$,
\begin{equation}
   z_L=\langle\Psi_0|
 \exp\left[{\rm i}\frac{2\pi}{L}\sum_{j=1}^L j S_j^z\right]|\Psi_0\rangle,
    \label{eqn:def_z}
\end{equation}
plays a role of an order parameter which detects VBS ground states {\it
directly}, and that it can be applied to determination of phase
boundaries in between different VBS states.  We demonstrate this idea in
the bond-alternating Heisenberg spin chains, and in the $S=1/2$ two-leg
frustrated ladder.

\section{Properties of the Order Parameter}\label{sec:properties}
\subsection{The valence-bond-solid picture}
According to the LSM argument, if the unique ground state
$|\Psi_0\rangle$ and $U|\Psi_0\rangle$ with $U\equiv\exp[(2\pi{\rm
i}/L)\sum_{j=1}^L j S_j^z]$ are orthogonal ($z_L=0$), there exists at
least one eigenstate with energy of ${\cal O}(L^{-1})$.  Now we consider
the overlap $z_L$ in the opposite situations such as the Haldane gap
phase.  In order to calculate $z_L$ explicitly in VBS states, we
introduce the Schwinger boson representation for the spin operators:
\begin{equation}
  S_j^{+}=a_j^{\dag}b_j,\ \ \ 
  S_j^{-}=b_j^{\dag}a_j,\ \ \ 
  S_j^{z}=\frac{1}{2}(a_j^{\dag}a_j-b_j^{\dag}b_j),\ \ \ 
  S_j=\frac{1}{2}(a_j^{\dag}a_j+b_j^{\dag}b_j),
\end{equation}
where these bosons satisfy the commutation relation
$[a_i,a_j^{\dag}]=[b_i,b_j^{\dag}]=\delta_{ij}$ with all the other
commutations vanishing.  $a_j^{\dag}$ ($b_j^{\dag}$) increases the
number of up (down) $S=1/2$ variables under symmetrization.  Then, a
generalized VBS state in a periodic system is written as
\begin{equation}
 |\Psi_{\rm VBS}^{(m,n)}\rangle
  \equiv\frac{1}{\sqrt{\cal N}}\prod_{k=1}^{L/2}
  (B_{2k-1,2k}^{\dag})^{m}(B_{2k,2k+1}^{\dag})^{n}|{\rm vac}\rangle,
  \label{eqn:VBS}
\end{equation}
where $B_{i,j}^{\dag}\equiv
a_{i}^{\dag}b_{j}^{\dag}-b_{i}^{\dag}a_{j}^{\dag}$, ${\cal N}$ is a
normalization factor, and $|{\rm vac}\rangle$ is the vacuum with respect
to bosons.  $m$ and $n$ are integers satisfying $m+n=2S$.  Using
relations $Ua_j^{\dag}U^{-1}=a_j^{\dag}{\rm e}^{+{\rm i}\pi j/L}$ and
$Ub_j^{\dag}U^{-1}=b_j^{\dag}{\rm e}^{-{\rm i}\pi j/L}$, a twisted
valence bond $UB_{j,j+1}^{\dag}U^{-1}$ for $1\leq j\leq L-1$, and that
located at the boundary are calculated as follows,
\begin{eqnarray}
 UB_{j,j+1}^{\dag}U^{-1}&=&
 {\rm e}^{-{\rm i}\pi/L}a_{j}^{\dag}b_{j+1}^{\dag}
 -{\rm e}^{ {\rm i}\pi/L}b_{j}^{\dag}a_{j+1}^{\dag},\label{eqn:twbond1}\\
 UB_{L,1}^{\dag}U^{-1}&=&
 -({\rm e}^{-{\rm i}\pi/L}a_{L}^{\dag}b_{1}^{\dag}
 -{\rm e}^{ {\rm i}\pi/L}b_{L}^{\dag}a_{1}^{\dag}).\label{eqn:twbond2}
\end{eqnarray}
In the latter case, a negative sign appears for each valence bond.  Thus
the asymptotic form of $z_L$ is given by
\begin{equation}
 z_L=\langle\Psi_{\rm VBS}^{(m,n)}|U|\Psi_{\rm VBS}^{(m,n)}\rangle
  =(-1)^n[1-{\cal O}(1/L)].\label{eqn:sign_relation}
\end{equation}
It turns out that $z_L$ changes its sign according to the number of
valence bonds at the boundary.  $z_L$ can be calculated in more detail
by the matrix product formalism.

\subsection{The sine-Gordon theory}
Next, we consider $z_L$ from the viewpoint of a low energy effective
theory. In general, the Lagrangian density of a spin chain is given by
the sine-Gordon model,
\begin{equation}
 {\cal L}=\frac{1}{2\pi K}\left[\nabla \phi(x,\tau)\right]^2
  -\frac{y_{\phi}}{2\pi\alpha^2}\cos[\sqrt{2}\phi(x,\tau)],
  \label{eqn:SG_model}
\end{equation}
where $\tau$ is the imaginary time, $\alpha$ is a short range cut off,
and $K$ and $y_{\phi}$ are the parameters determined phenomenologically.
In the gapped (gapless) region one has $y_{\phi}(l)\rightarrow\pm\infty$
($y_{\phi}(l)\rightarrow 0$) for $l\rightarrow\infty$ under
renormalization $\alpha\rightarrow{\rm e}^{l}\alpha$.  On the unstable
Gaussian fixed line [$y_{\phi}(0)=0$ with $K(0)<4$], a second-order
``Gaussian transition'' takes place between the two gapped states.  In
this formalism, the spin wave excitation created by $U$ corresponds to
the vertex operator $\exp({\rm i}\sqrt{2}\phi)$, so that $z_L$ is
related to the ground-state expectation value of the nonlinear term as
$z_L\propto\langle\cos(\sqrt{2}\phi)\rangle$ and the three fixed points
$y_{\phi}=\pm\infty,0$ correspond to $z_{\infty}=\mp 1,0$,
respectively. Thus the Gaussian critical point can be identified by
observing $z_{L}=0$ \cite{Nakamura-V}.

\section{Application to Physical Systems}
\subsection{The bond-alternating Heisenberg chain}\label{sec:BAHC}
In the bond-alternating Heisenberg chain,
\begin{equation}
  {\cal H}=\sum_{j=1}^L[1-\delta(-1)^{j}]\bm{S}_{j}\cdot\bm{S}_{j+1},
   \label{eqn:BAHC}
\end{equation}
the VBS picture is considered to be realized approximately: The
configuration of the valence bonds $(m,n)$ changes from $(0,2S)$ to
$(2S,0)$ successively as $\delta$ is increased form $-1$ to $1$, meaning
the existence of $2S$ quantum phase transitions.  We calculate $z_L$ for
$2S=1,2,3,4$ in a periodic system by the quantum Monte Carlo (QMC)
method with the continuous-time loop algorithm \cite{Todo-K}, and
confirmed that qualitative behavior of $z_L$ agrees with our theory
given in Sec.~\ref{sec:properties}.

Next, we determine the successive dimerization transition points by
observing $z_L=0$ with $L$ up to $320$.  Extrapolation of the data has
been done by the function $\delta_{\rm c}(L)=\delta_{\rm
c}(\infty)+A/L^2+B/L^4+C/L^6$.  For $S=1$ [$(1,1)$-$(2,0)$], $S=3/2$,
[$(2,1)$-$(3,0)$] and $S=2$ [$(2,2)$-$(3,1)$, $(3,1)$-$(4,0)$] cases, we
obtain the transition points $\delta_{\rm c}=0.25997(3), 0.43131(7),
0.1866(7), 0.5500(1)$, respectively, where $(\,)$ denotes 2$\sigma$.
The results are consistent with the previous estimates, but much more
accurate: $\delta_{\rm c}=0.2595(5)$ for the $S=1$ case by the QMC
calculation for the susceptibility \cite{Kohno-T-H}, and $\delta_{\rm
c}=0.2598, 0.4315, 0.1830, 0.5505$ obtained by the level-crossing
method~\cite{Kitazawa-N}.

\subsection{The spin-$1/2$ frustrated ladder}\label{sec:ladder}
We discuss $z_L$ in the $S=1/2$ two-leg ladder with frustration:
\begin{equation}
 {\cal H}=
  \sum_{j=1}^L[{\bm S}_{j}\cdot{\bm S}_{j+1}
  +{\bm T}_{j}\cdot{\bm T}_{j+1}
  +J_{\bot}{\bm S}_{j}\cdot{\bm T}_{j}
  +J_{\times} ({\bm S}_{j}\cdot{\bm T}_{j+1}
  +{\bm T}_{j}\cdot{\bm S}_{j+1})],
\end{equation}
where ${\bm S}_{j}$ and ${\bm T}_{j}$ denote $S=1/2$ operators at $j$-th
site. In has been pointed out~\cite{Kim-F-S-S} that there appears
competition between a rung dimer phase ($J_{\times}\ll J_{\bot}$) and a
Haldane phase ($J_{\bot}\ll J_{\times}\leq 1$), and that these two
phases are identified by two distinct SOPs, ${\cal O}_{\rm
odd}^{\alpha},{\cal O}_{\rm even}^{\alpha}$ given by
Eq.(\ref{eqn:string}) where $S_j^{\alpha}$ is replaced by the following
two types of composite spin operators:
\begin{equation}
 \tilde{S}^\alpha_{{\rm odd},j}
  \equiv S^{\alpha}_{j} + T^{\alpha}_{j},\ \ \ 
 \tilde{S}^\alpha_{{\rm even},j}
 \equiv S^{\alpha}_{j} + T^{\alpha}_{j+1}.
\end{equation}
Then, ${\cal O}_{\rm odd}=0$ and ${\cal O}_{\rm even}\neq0$ for the rung
dimer phase, and ${\cal O}_{\rm odd}\neq 0$ and ${\cal O}_{\rm even}=0$
for the Haldane phase.  Instead of ${\cal O}_{\rm odd}$ and ${\cal
O}_{\rm even}$, we introduce $z_{\rm odd}$ and $z_{\rm even}$ defined
\begin{wrapfigure}{r}{7.0cm}
 \epsfxsize=7.0cm \epsfbox{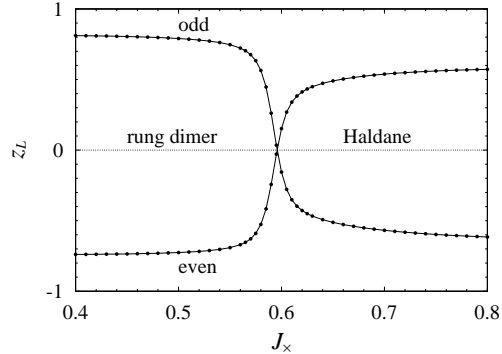}
 \caption{$J_{\times}$-dependence of $z_{\rm odd}$ and $z_{\rm even}$ in
 the $S=1/2$ frustrated ladder for $J_{\bot}=1$ and $L=14$, calculated
 by the exact diagonalization. The point $z_{\rm odd}=z_{\rm even}=0$
 gives the dimer-Haldane phase boundary.}\label{fig:1}
\end{wrapfigure}
by Eq.(\ref{eqn:def_z}) where $S_j^{z}$ is replaced by
$\tilde{S}^z_{{\rm odd},j}$ and $\tilde{S}^z_{{\rm even},j}$.  In
Fig.~\ref{fig:1}, we show $z_{\rm odd}$ and $z_{\rm even}$ for
$J_{\bot}=1$ and $L=14$ calculated by the exact diagonalization.  We
find that $|z_{\rm odd}|\neq|z_{\rm even}|$ in general, and $z_{\rm
odd}$ and $z_{\rm even}$ have opposite signs.  Especially, $z_{\rm
odd}=z_{\rm even}=0$ at the same point which corresponds to the
dimer-Haldane phase boundary. This tells us that one of the two $z_L$
has enough information to describe this system, since $z_L$
characterizes the system by its magnitude and sign, while the SOPs give
zero or finite value without changing the sign. Thus $z_L$ turns out to
be a more rational order parameter to describe VBS states in a unified
way.  The properties of $z_L$ is also useful to determine the accurate
phase boundary from numerical data of finite-size clusters.

\section{Summary}
We have introduced $z_L$ given by Eq.~(\ref{eqn:def_z}) as an order
parameter to characterize VBS states.  The proposed order parameter
changes its sign according to the configuration of valence bonds.  This
property enables us to determine the critical point between different
VBS states by observing $z_L=0$.  We have demonstrated this theory for
phase transitions in the bond-alternating Heisenberg chain, and in the
frustrated spin ladder.  Details of the present results will be
published elsewhere \cite{Nakamura-T}.

M.N. thanks M. Oshikawa and J. Voit for useful discussions, and Y. Oi
Nakamura for encouragement.  S.T. acknowledges support of the Swiss
National Science Foundation.  The QMC simulations were performed on SGI
2800 at the Supercomputer Center, Institute for Solid State Physics,
University of Tokyo.


\end{document}